\documentclass{iopart}
\usepackage{graphicx}
\usepackage{amssymb}
\begin{document}

\title{Polarized Spots in  Anisotropic Open Universes}

\author{Rockhee Sung and Peter Coles} \address{School of Physics \& Astronomy, Cardiff University, 5 The Parade,
Cardiff CF24 3AA, UK} \ead{Peter.Coles@astro.cf.ac.uk}

\begin{abstract}
We calculate the temperature and polarization patterns generated in
anisotropic cosmological models drawn from the Bianchi
classification. We show that localized features in the temperature
pattern, perhaps similar to the cold spot observed in the Wilkinson
Microwave Anisotropy Probe (WMAP) data, can be generated in models
with negative spatial curvature, i.e. Bianchi types V and VII$_{h}$.
Both these models also generate coherent polarization patterns. In
Bianchi VII$_h$, however, rotation of the polarization angle as
light propagates along geodesics can convert E modes into B modes
but in Bianchi V this is not necessarily the case. It is therefore
possible, at least in principle, to generate localized temperature
features without violating existing observational constraints on the
odd-parity component of the cosmic microwave background
polarization.
\end{abstract}
 \submitto{\CQG}

\section{Introduction}
\label{sec:intro}

Observations of the temperature anisotropies of the cosmic microwave
background, particularly those from the Wilkinson Microwave
Anisotropy Probe (WMAP) \cite{WMAP1,WMAP2}, form the foundations of
the current (``concordance'') cosmological model \cite{Coles}.
However, WMAP has also uncovered tantalizing evidence of departures
from the standard framework. In particular, detailed analysis of the
pattern of temperature fluctuations has led to the identification of
a \emph{Cold Spot}
\cite{Vielva,Cruz1,Cruz2,Cruz3,Cruz4,Cayon,Cspot}. The level of
departure from isotropy is relatively small but it is highly
significant from a statistical point of view. The presence of this
feature seems to be inconsistent with the assumption of global
isotropy upon which the concordance cosmology is based and could be
evidence that we live in a globally anisotropic Universe, i.e. one
not described by a Friedmann-Robertson-Walker (FRW) model.

The Bianchi classification arranges all possible spatially
homogeneous but anisotropic relativistic cosmological models into
types depending on the symmetry properties of their spatial
hypersurfaces \cite{Gris,Ellis1}. It has been known for some time
that localized features in the radiation background can occur in
Bianchi models with negative spatial curvature
\cite{Collins1,Dautcourt1,Tolman1,Tolman2,Barrow1}. The physical
origin of such features lies in the focussing effect of spatial
curvature on the geodesics that squeezes the pattern of the
anisotropic radiation field into a small region of the sky. Only a
few of the Bianchi types contain the FRW model as a limiting case
and, from this subset the model which appears to best able to
reproduce the anomalous cold spot is the Bianchi VII$_h$ case
\cite{Jaffe1,Jaffe2,McEwan1,McEwan2,Bridge}. However, as well as
forming distinctive features in the temperature pattern, anisotropic
cosmological models also generate characteristic signatures in the
polarized component of the background radiation. Thomson scattering
generates polarization as long as there is a quadrupole anisotropy
in the temperature field of the radiation incident upon the
scattering particle. In the concordance cosmology the temperature
and polarization patterns are (correlated) stochastic fields arising
from their common source in scalar and tensor perturbations arising
from inflation. In a Bianchi cosmology, however, the patterns are
coherent and have a deterministic relationship to one another owing
to their common geometric origin. It has recently been shown
\cite{Pontzen1,Pontzen2,Sung1} that the properties of the
polarization field produced in Bianchi VII$_h$ are inconsistent with
the latest available WMAP polarization data \cite{WMAPPol} because
they inevitably involve a large odd-parity (B-mode) contribution
that exceeds the experimental upper limit.

In a forthcoming paper we present an exhaustive study of the
temperature and polarization anisotropies produced by those Bianchi
types that possess an FRW limit \cite{Sung1}. The purpose of
studying these models is to characterize as fully as possible the
radiation fields they can produce in order to separate them as
clearly as possible from residual systematics and thus provide the
strongest possible constraints on exotic cosmologies. In the present
paper, however, we focus on the very specific question of whether
localized spots necessarily involve a large B-mode polarization.

\section{Bianchi Cosmologies}
\label{sec:Bianchi}


The models we consider are based on Einstein's general theory of
relativity and we use the field equations in the form
\begin{equation}
G_{ab}\equiv R_{ab}-\frac{1}{2} R g_{ab} = T_{ab}-\Lambda g_{ab},
\end{equation}
with $R_{ab}$ being the Ricci Tensor, $R$ the Ricci scalar, $T_{ab}$
the energy-momentum tensor and $\Lambda$ the cosmological constant.
Indices $a$ and $b$ run from $0$ to $3$. We use units where $8\pi
G=c=1$. In terms of a coordinate system $x^{a}$, the metric $g_{ab}$
is written
\begin{equation}
ds^2=g_{ab} dx^{a}dx^{b}=(h_{ab}-u_{a}u_{b}) dx^{a}dx^{b},
\end{equation}
where $u^{a}$ is the fluid velocity; the signature of $g_{ab}$ is
$(-+++)$. Starting from a local coordinate system $x^{i}$ we
construct a tetrad basis \cite{Ellis1}
\begin{equation}  {\bf e}_a= e_{a}^{i} \frac{\partial}{\partial
x^{i}}\label{tet1}
\end{equation}
such that \begin{equation} g_{ab}=e_{a}^i e_{b}^j
g_{ij}=e_{a}^{i}e_{bi} = {\rm diag} (-1, +1, +1, +1)\end{equation}
meaning that the tetrad basis ${\bf e}_a$ is orthonormal. The Ricci
rotation coefficients, \begin{equation}
\Gamma_{abc}=e_{a}^{i}e_{ci;j} e_b^{j}, \end{equation} are the
tetrad components of the Christoffel symbols; semicolons denote
covariant derivatives. In general, the operators defined by equation
(\ref{tet1}) do not commute: they generate a set of relations of the
form
\begin{equation} [{\bf e}_a, {\bf e}_b]=\gamma_{ab}^{c} {\bf
e}_{c}\label{comm}. \end{equation} The Ricci rotation coefficients
are just \begin{equation} \Gamma_{abc}= \frac{1}{2}
(\gamma_{abc}+\gamma_{cab}-\gamma_{bca}).
\end{equation} The matter flow is described in terms of the expansion
$\vartheta_{ab}$ and shear $\sigma_{ab}$:
\begin{eqnarray}
u_{a;b}& =& \omega_{ab}+\vartheta_{ab}-\dot{u}_au_{b}\nonumber\\
\sigma_{ab}& = & \vartheta_{ab}-\frac{1}{3} \vartheta h_{ab}\equiv
\vartheta_{ab}-Hh_{ab},
\end{eqnarray}
where $\vartheta=\Tr(\vartheta_{ab})=\vartheta_{aa}$ and the
magnitude of the shear is $\sigma^2=\sigma^{ab}\sigma_{ab}/2$.
 We now take the time-like vector in our basis to be the fluid flow
velocity so that $u^a=\delta_{0}^a$ and $u_a=-\delta_{a}^0$. The
remaining space-like vectors  form an orthonormal triad, with a set
of commutation relations like that shown in equation (\ref{comm})
but with an explicit time dependence in the ``structure constants''
describing the spatial sections:
\begin{equation}
[{\bf e}_i, {\bf e}_j]=\gamma_{ij}^{k}(t) {\bf e}_{k}.
\end{equation}
Without loss of generality we can write \begin{equation}
\gamma_{ij}^{k}=\epsilon_{ijl}n^{lk}+\delta_{j}^{k}a_{i}-\delta_i^{k}a_{j},
\end{equation}
for some tensor $n_{ij}$ and some vector $a_i$. The Jacobi
identities require that $n_{ij}a^{j}=0$ so we choose $a^{j}=(a,0,0)$
and $n_{ij}={\rm diag}(n_1,n_2,n_3)$. The four remaining free
parameters are used to construct the Bianchi classification
described in detail elsewhere \cite{Gris,Ellis1}.

We are interested in cosmological models that are close to the
homogeneous and isotropic FRW case, but not all the Bianchi types
contain this solution. Those  that do are types I, V, VII$_{0}$,
VII$_{h}$ and IX. Bianchi I and Bianchi VII$_{0}$ are spatially
flat, Bianchi IX is positively curved and Bianchi types V and
VII$_{h}$ have negative spatial curvature. The open cases are of
particular interest in this paper as they permit the focussing of
anisotropic patterns into small regions of the sky. The scalar
curvature $R$ of the spatial sections is given in terms of the
Bianchi parameters as
\begin{equation}
R=\frac{1}{2}(2n_1n_2+2n_1n_3+2n_2n_3-n^2_1-n^2_2-n^2_3)-6a^2.
\end{equation}
For Bianchi V we have $n_1=n_2=n_3=0$ so that $R=-6a^2$. In Bianchi
VII$_h$ we have $n_1=0$ but $n_2\neq 0$ and $n_3\neq 0$; the
parameter $h$ is defined by $h=a^2/n_2n_3$. If we take
$n_{2}=n_3=n$, the canonical form of the model, we also have
$R=-6a^2=-6n^{2}h$.

The models we consider have a single preferred axis of symmetry. The
alignment of the shear eigenvectors relative to this preferred axis
determines not only the dynamical evolution of the model through the
field equations, but also the temperature and polarization pattern
that gets imprinted into the cosmic background radiation.

\section{Temperature and Polarization Anisotropies}

The transfer equation for polarized radiation propagating through
space-time can be described by a combination of propagation vector
and a polarization vector. In the tetrad frame these can be written
as
\begin{equation}\label{photon_distribution}
\hat{N} \equiv {N^0 \choose N^2} = \frac{1}{ch^4\nu^3} {I+iV \choose
Q-iU},
\end{equation}
in terms of the standard Stokes parameters, $I$, $Q$, $U$ and $V$
where the superscripts indicate the parts of the (complex) photon
distribution involving  spin weights $0$ and $2$. The direction of a
light ray in three-dimensional space is defined by  $k^i=
(\cos\theta,\sin\theta \cos\phi, \sin\theta \sin\phi )$ and we have
also used the complex unit vector
\begin{equation} m^i=-\frac{1}{\sqrt{2}}\eth
k^i \equiv -\frac{1}{\sqrt{2}}(\frac{\partial}{\partial \theta}
+\frac{i}{\sin \theta}\cdot\frac{\partial }{\partial \phi})k^i.
\end{equation}
Neglecting shear-dependent terms in the geodesic equation,
the following  equation can be obtained
\cite{Dautcourt1,Tolman1} that describes the evolution of the
direction of propagation:
\begin{eqnarray}
\label{rotate0} \frac{1}{\varepsilon}\frac{d k_i}{dt} & = &
 \epsilon_{ijk}n^{jl}k_lk^k+a_jk_ik^j-a_i,
\end{eqnarray}
$\varepsilon$ is the photon energy. This leads to
\begin{eqnarray}
\label{rotate1} \frac{1}{\varepsilon}\frac{d\theta}{dt} & = &
\left[a+(n_3-n_2)\cos\phi\sin\theta\right]\sin\theta \nonumber\\
\frac{1}{\varepsilon}\frac{d\phi}{dt} & = & \left[n_1-n_3 +(n_3-n_2)
\cos^2 \phi\right] \cos\theta.
\end{eqnarray}
The radiation distribution described by $N^{A}$ is affected by the
properties of the geodesics along which photons propagate but is
also altered by scattering. If only elastic scattering is included,
this is governed by a Boltzmann equation
\begin{equation}
\label{Boltza}
 {\cal D} N^A  = \tau(-N^A + J^A),
\end{equation}
involving the operator
\begin{equation}
 \label{Boltz}{\cal D} \equiv \frac{\partial}{\partial t}-\varepsilon\gamma^0\frac{\partial}{\partial\varepsilon}
    + \frac{\gamma^i}{\sqrt{2}}
    (m^i\bar{\eth}+\bar{m}^i\eth)  +
    i\delta^2_A\Gamma^k_{\phantom{0}li}k^lk^m\epsilon^{ikm},
    \end{equation}
where $\tau$ the optical depth; the amplitude of the polarization
depends strongly on $\tau$ but we shall not explore its dependence
in detail in this paper. The horizontal bar denotes complex
conjugation. We also have $ \gamma^a = \Gamma^a_{\phantom{0}i0}k^i+
\Gamma^a_{\phantom{0}ij}k^ik^j $, so the non-zero connection terms
are given by
\begin{eqnarray}
  \Gamma^0_{\phantom{0}ij} &= \Gamma^i_{\phantom{0}j0} &= \vartheta_{ij}  \nonumber\\
  \Gamma^i_{\phantom{0}jk} &= -\Gamma^k_{\phantom{0}ji} &
  =\frac{1}{2}(\epsilon_{ljk}n^l_i-\epsilon_{ilk}n^l_j+\epsilon_{ijl}n^l_k)+(\delta^k_ja_i-\delta^i_ja_k)
\end{eqnarray}
if we assume no vorticity ($\omega^k =0$) and no acceleration
($\Gamma^0_{\phantom{0}0i} =\dot{u}_i $). The Boltzmann operator
(\ref{Boltz}) involves terms that describe the propagation of the
radiation field along the geodesics, the effect of the shear on the
photon energy distribution, and the rotation of the polarization
angle $\chi$ defined by $2\chi= \arctan (U/Q)$. The emission term
$J_A$ (which involves terms describing the fluid flow) contains only
harmonics up to $l=2$, since the radiation modes with $l\leq2$ are
damped as well as re-radiated by Thomson scattering, while
higher-order modes $l>2$ are only damped.

\begin{figure}[htp]
  \begin{center}
      \includegraphics[width=0.2\textwidth,angle=90]{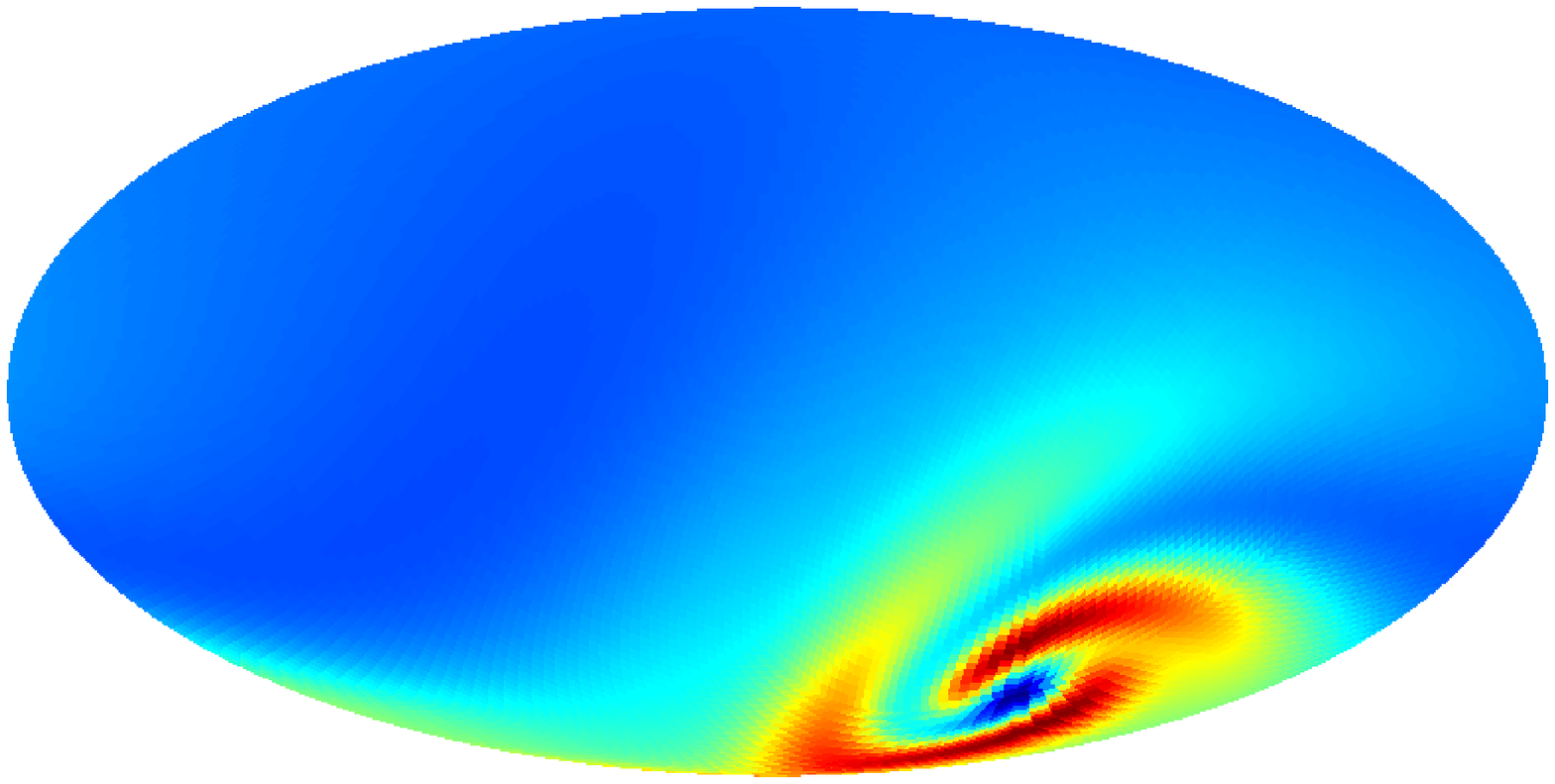}
   \includegraphics[width=0.2\textwidth, angle=90]{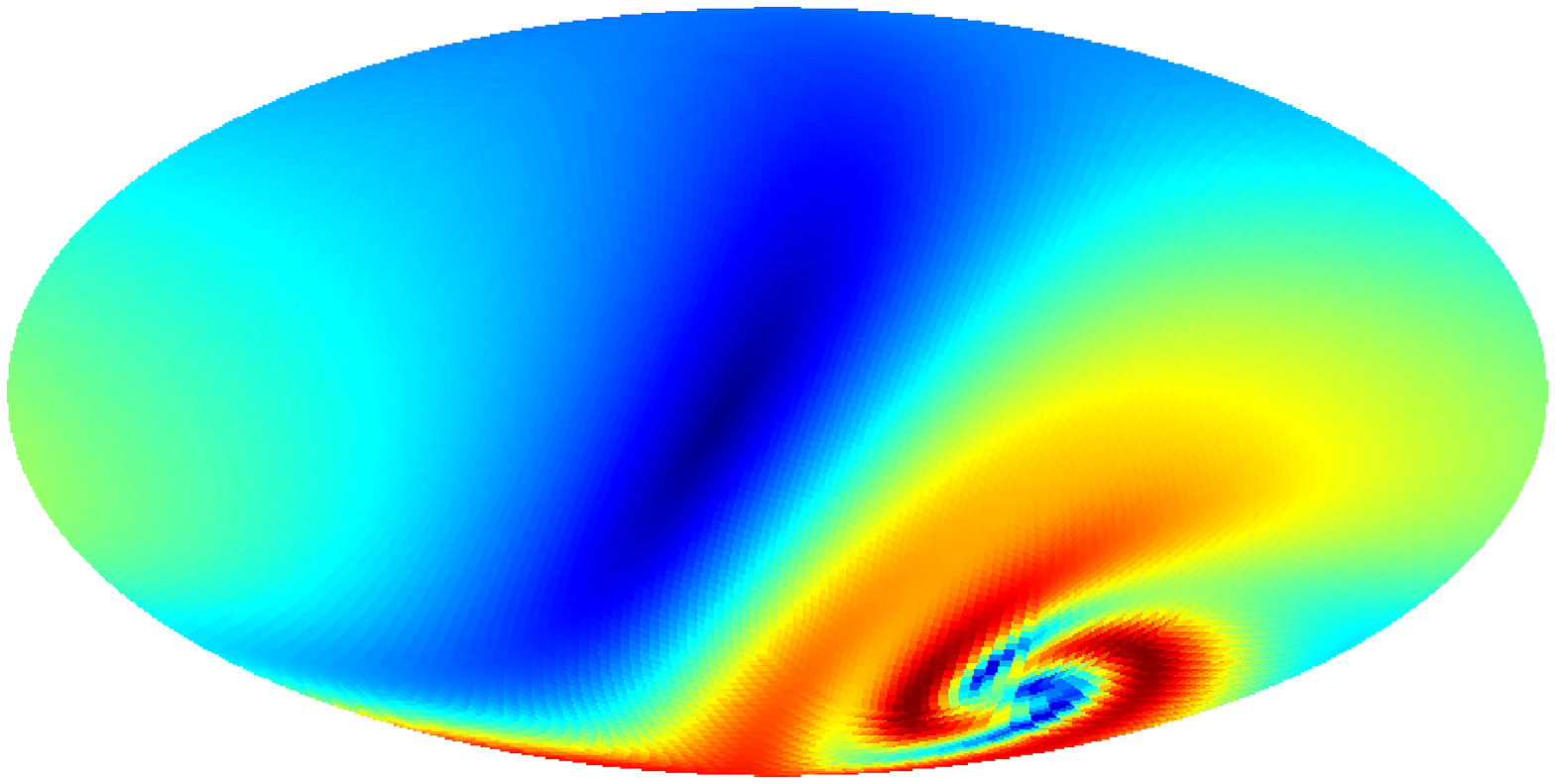}\\
    \includegraphics[width=0.2\textwidth,angle=90]{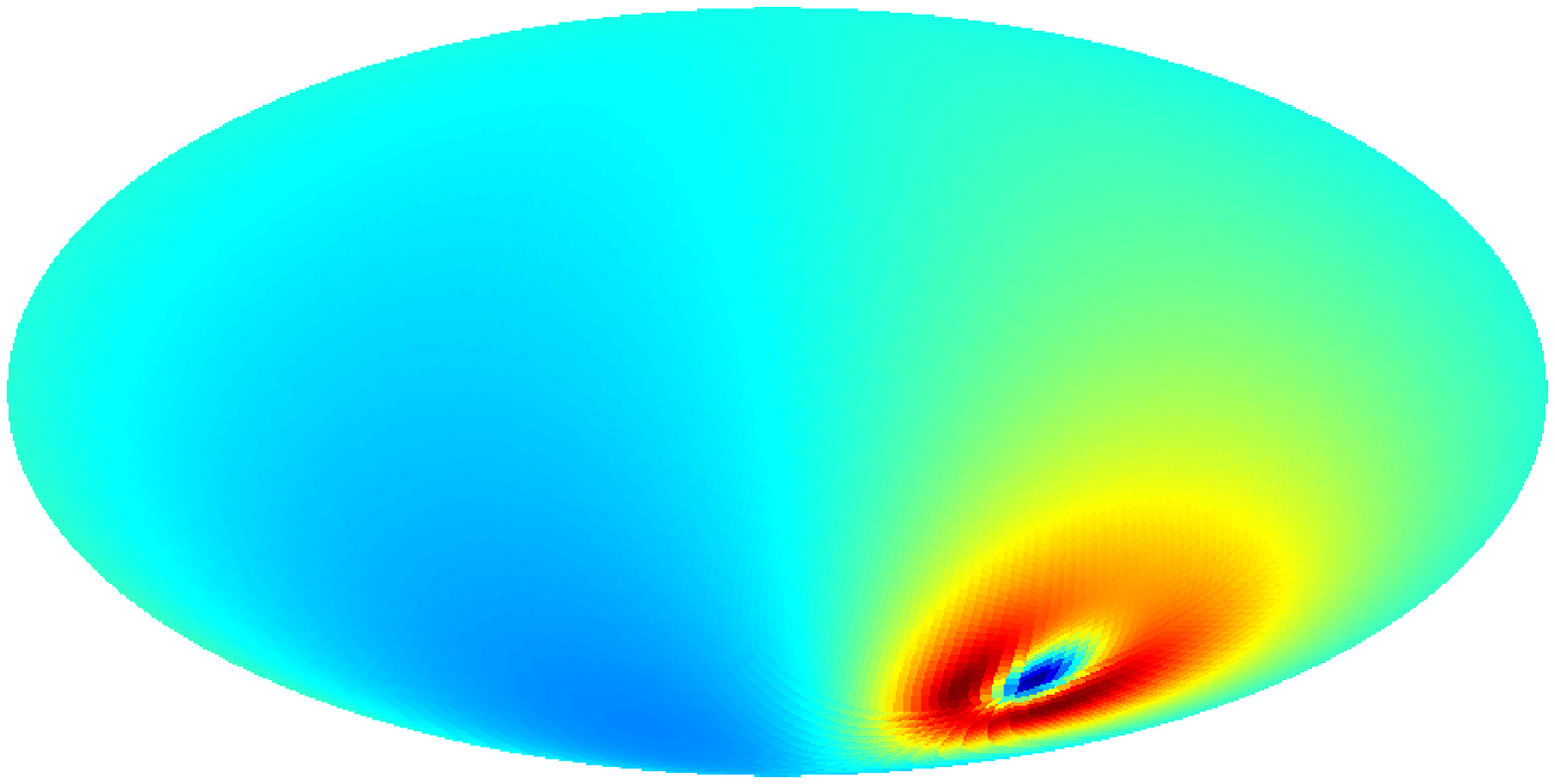}
   \includegraphics[width=0.2\textwidth, angle=90]{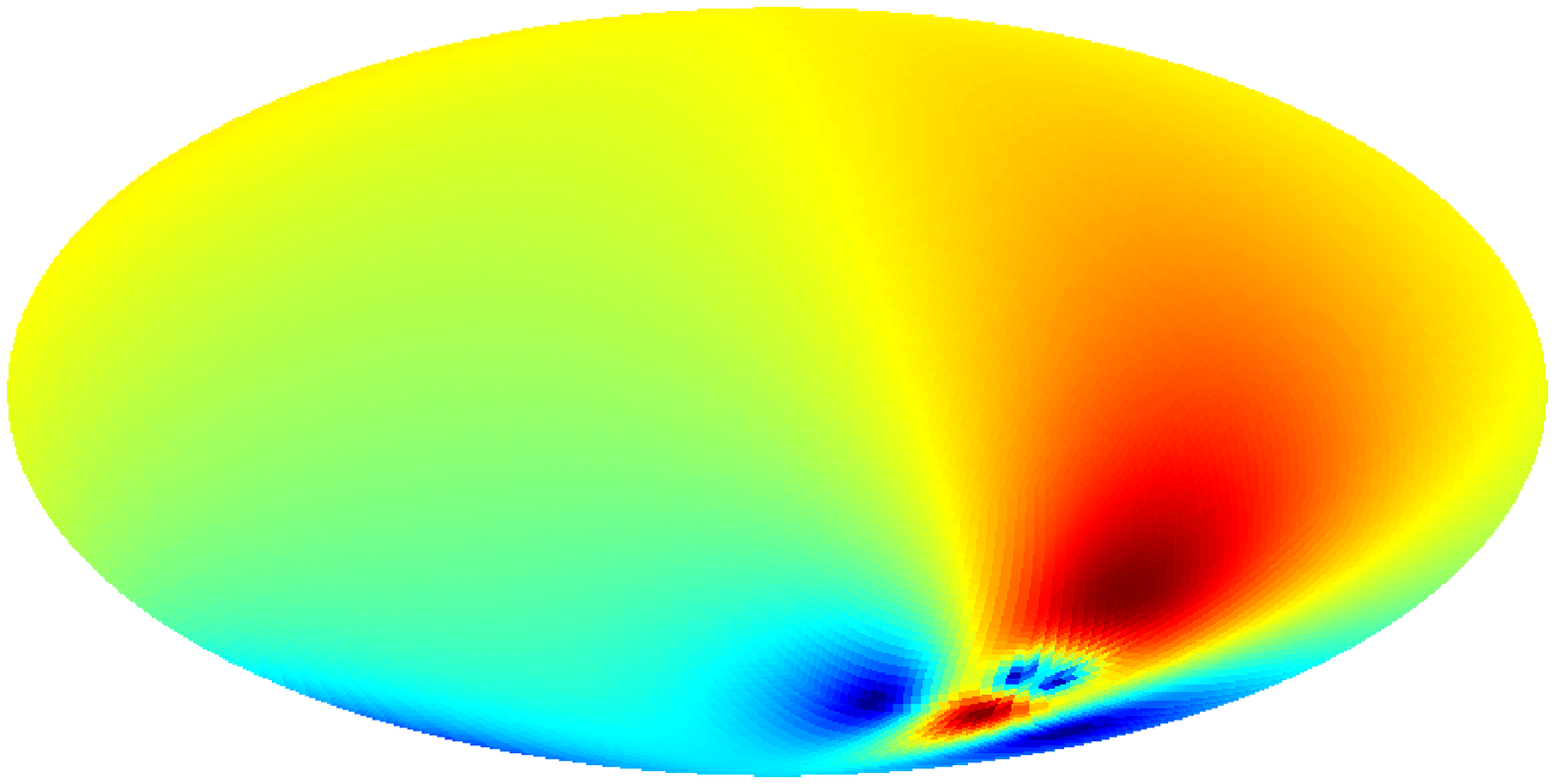}
   \end{center}
\caption{Temperature and Polarization maps for Bianchi models with
localized features (aligned with the known CMB Cold Spot). The left
panels show the temperature, the right the total polarization
$P=\sqrt{Q^2+U^2}$. The top row shows  an example of Bianchi VII$_h$
with a compact spiral feature produced by focussing and twisting;
the polarized component contains a significant B-mode. The bottom
row shows a Bianchi V model in which the initial quadrupole is
focussed but not twisted. The polarized field in the latter case has
no B-mode.}
\end{figure}
It has become conventional to decompose the polarized component of
the cosmic microwave background into modes classified according to
their parity. The even modes are called E-modes and the odd modes
are the B-modes. The latter are of particular interest in the
context of inflationary cosmology as they cannot be sourced by
scalar perturbations and are therefore generally supposed to be a
signature of the presence of primordial tensor perturbations, i.e.
gravitational waves \cite{Kam,Hu1}
\begin{eqnarray*}
N^2 &\equiv& N^2_{ij}m^{ij}= (Q-iU)(\hat{n})=\sum_{lm}a_{-2,lm} {\phantom{a}}_{-2} Y_{lm}(\hat{n}),\\
\bar{N}^2 &\equiv& \overline{N^2_{ij}m^{ij}}=
(Q+iU)(\hat{n})=\sum_{lm}a_{2,lm} {\phantom{a}}_2 Y_{lm}(\hat{n})
\end{eqnarray*}
\begin{equation}
a_{E,lm}=-(a_{2,lm}+a_{-2,lm})/2, \qquad
a_{B,lm}=i(a_{2,lm}-a_{-2,lm})/2
\end{equation}
The overall {\em level} of polarization (in both E and B modes)
increases strongly with the optical depth to Thomson scattering
$\tau$ through (\ref{Boltza}).

However, the equations of the previous section for the evolution of
the polarized radiation distribution allow us to establish some firm
implications for the relative size E and B modes just by considering
their symmetry. First, the scattering term $J^A$ produces a pure
E-mode quadrupole anisotropy. However, depending on the initial
conditions, the redshifting effect of shear can produce either E or
B modes (just as a gravitational wave perturbation of FRW can). The
physical processes described by Equations (14)-(17) can, in
principle, convert E modes into B modes and vice versa. In Bianchi
VII$_h$ these effects are {\em unavoidable} so, even if there is no
initial B-mode, one is inevitably generated as the Universe evolves.
In Bianchi V, however, the last term in Equation (17) does not
contribute to the mixing of E and B modes, at least at this level of
perturbation theory. Since initial conditions exist in which the
polarization is purely E mode there are therefore models of this
type that can produce spots without any B modes. Figure 1, which was
plotted using the Healpix software \cite{healpix}, shows
illustrative examples of Bianchi VII$_h$ and Bianchi V that show
localized patterns with and without B-modes respectively.

\section{Discussion and Conclusions}
\label{sec:disc}

We have shown that it is possible, in a Bianchi V cosmological
model, to generate a localized temperature anomaly qualitatively
similar to the known Cold Spot without necessarily producing a large
B-mode polarization. Observational limits on the B mode alone are
therefore not sufficient to exclude global anisotropy as a possible
explanation of the famous Cold Spot. Although the set of Bianchi
models that can evade the limits set by B mode polarization is
small, a more rigorous analysis requires a fuller parametrization in
which the Bianchi parameters are constrained with other cosmological
parameters (such as the optical depth) using a complete description
of the polarized radiation field.

We stress that there are severe difficulties with anisotropic
universes as explanations for the overall pattern of observed CMB
anomalies. Most important among these is that a significant
(negative) spatial curvature seems to be at odds with measurements
that clearly prefer a flat universe \cite{WMAP1,WMAP2}.
Nevertheless, these models may provide important clues that can lead
to more effective and efficient use of observations to test exotic
cosmologies. For example, there are other ways in which the
polarization angle for CMB photons can be rotated \cite{Ni}; the
lack of any observed B-mode allows present observations to place
strong constraints on such models also \cite{Wu}.

\section*{Acknowledgments}

We thank Jason McEwen, Sasha Polnarev and Leonid Grishchuk for
useful discussions. Rockhee Sung acknowledges an Overseas
Scholarship from the Korean government.

\end{document}